\documentclass[cameraready]{Interspeech}

\usepackage{graphicx}
\usepackage{amsmath}
\usepackage{xurl}

\title{Your Home Remembers: A Voice-Enabled Household Memory Assistant \\ for Conversational Retrieval}

\author[affiliation={1,2}, orcid=0009-0007-7259-656X]{Aleix}{Sant}
\author[affiliation={1}, orcid=0000-0002-4507-4930]{Jordi}{Luque}
\author[affiliation={1, 3}, orcid=0000-0002-7705-2250]{Fernando}{López}
\author[affiliation={1}]{Pablo}{Gómez}
\author[affiliation={2}, orcid=0000-0001-6657-673X]{Carlos}{Escolano}

\address{
    $^1$ Telefónica Innovación Digital, Scientific Research \\
    $^2$ Universitat Politècnica de Catalunya \\
    $^3$ Universidad Autónoma de Madrid
}
\email{aleix.santsavall@telefonica.com}


\begin{document}
\maketitle

\begin{abstract}
Everyday household information is often exchanged in conversations but never written down, making details easy to forget or misremember. We present a voice-enabled smart-home assistant that acts as a household memory by recording conversations and retrieving information from stored and newly captured speech. The assistant runs on an Android tablet with a modular architecture in which speech-recognition, diarisation, embedding and language models can be configured or replaced through an orchestration layer. On-device wake-word detection and VAD-based query capture support hands-free interaction. Recorded audio is processed in configurable chunks, transcribed and diarised, converted into turn-aware passages with conversation-level timestamps, and indexed in a vector store. For memory queries, a retrieval-augmented generation pipeline selects key passages and uses them to generate an answer. The interface supports recording, conversation management and inspection of timestamped sources.
\end{abstract}

\section{Introduction}
Recovering a specific detail from an earlier household conversation can be difficult, particularly when several people, topics and decisions are involved. A relative may explain how medication should be taken, family members may agree on an appointment, or someone may mention where an important object has been stored. We present a voice-enabled assistant that transforms such interactions into a searchable conversational memory. Unlike conventional assistants that primarily answer from parametric knowledge, the proposed system grounds its responses in conversations explicitly recorded and selected by the household through retrieval-augmented generation (RAG)~\cite{lewis2020rag}.

For example, a motivating scenario involves a physician discussing a diagnosis, symptoms, treatment instructions, medication and contact details with an older adult and a family member. At a later time, the older adult may need to recover a specific detail from that exchange, such as the prescribed dosage, a future medical appointment or when to contact the medical service. The assistant is intended to help users retrieve what was actually said. It is not designed to replace a clinician, interpret symptoms or provide new medical advice. This scenario highlights the importance of preserving speaker attribution and temporal provenance, and of clearly distinguishing retrieved conversational evidence from model-generated wording.

The assistant records consented conversations and makes both stored interactions and newly captured speech searchable shortly after transcription. Its main contributions are: (i) retrieval from both long-term conversational memory and speech processed only moments earlier; (ii) a modular orchestration architecture that decouples the Android application from the speech and language models, allowing compatible models to be selected or integrated without changing the interaction logic; (iii) a turn-aware ingestion step that merges diarised segments into speaker turns and associates each indexed passage with conversation-level timestamps; and (iv) a tablet interface that supports recording and conversation management, as well as the inspection of timestamped passages used as evidence for each answer.

\section{Assistant interaction}
From the user's perspective, the assistant supports two connected activities: creating and managing conversational memories, and querying the selected memory through spoken interaction. The Android interface reflects this separation. To create a memory, the user names a conversation, selects the audio-chunk duration and presses \emph{Record Conversation}. A live waveform visualises the incoming audio level, while status messages make the capture and processing stages visible. During recording, the same control becomes \emph{Stop Conversation}, allowing the user to end audio capture and trigger ingestion of the final chunk explicitly. Stored conversations can be selected as the active retrieval context or permanently removed, while additional conversational memories can be created through new recordings. The interface also maintains a short-term history of the current question-answer exchange so that follow-up queries can use previous dialogue context. A \emph{Clear History} button removes only this short-term interaction history; it does not delete the selected recorded conversation or its indexed passages.

To enable hands-free question answering, the system is activated via a predefined wake-up word. When a conversation is selected, the assistant captures the request using Silero VAD~\cite{silero_vad}, sends it to the configured ASR service, retrieves relevant memory passages, and displays the generated answer. A source panel lists the retrieved text with labels such as ``Snippet 1 -- Time 00:12'', enabling the user to inspect the supporting evidence. The current prototype offers bilingual support for Spanish and English, configurable via a language selection menu.

The configuration dialogue exposes parameters for adapting the assistant's behaviour and retrieval pipeline, including the number of retrieved passages, response length, sampling settings, active language, embedding model, passage size and splitting strategy. Users can also specify the language and ASR models, provided that the models are available through the remote backend. These controls make it possible to balance response latency, retrieval precision, answer length and computational requirements without modifying the Android application.

\begin{figure*}[t]
  \centering
  \includegraphics[width=0.99\textwidth]{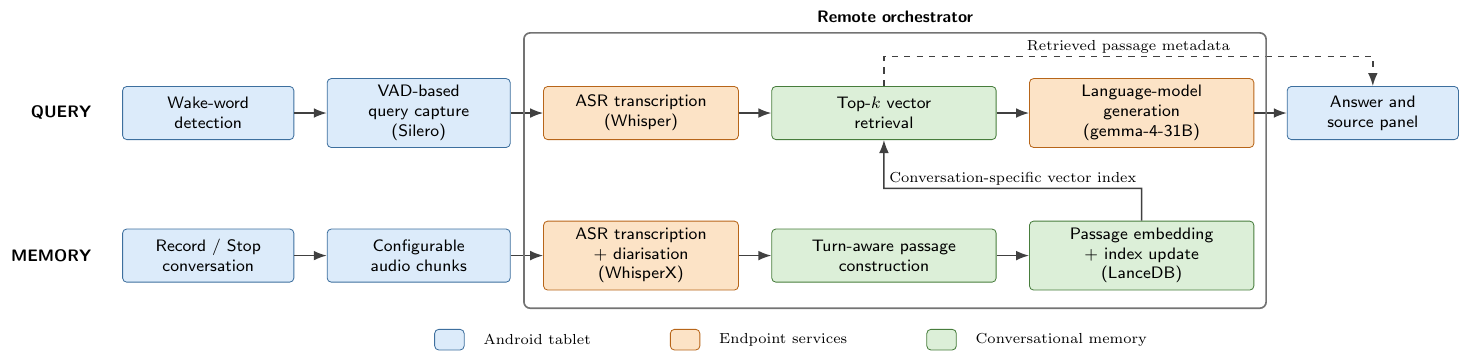}
  \caption{Assistant architecture. Blue blocks run on the tablet, orange blocks are model endpoint services and green blocks implement conversational memory. Solid arrows indicate processing and data flow, while the dashed path returns passage metadata to the interface without including it in the language-model prompt. The remote orchestrator routes application requests to the configured model and memory services through a common backend interface.}
  \label{fig:architecture}
\end{figure*}


\section{Technical architecture}
Figure~\ref{fig:architecture} presents the distributed architecture supporting the recording and question-answering workflows described above. The system comprises two coordinated processing paths. The query path performs wake-word detection and VAD-based capture on the tablet before remote transcription, retrieval and answer generation. The memory path records longer conversations in configurable audio chunks, transforms their diarised transcripts into turn-aware passages, and incrementally updates a conversation-specific vector index. This separation keeps latency-sensitive interaction on the device while allowing the speech, retrieval and language components to be configured and deployed as independent services.

\subsection{Orchestration and modular architecture}
The Android application\footnote{Android app source code: \url{https://github.com/Telefonica-Scientific-Research/eloquence_pilot1}} communicates with an orchestrator rather than directly with individual model endpoints. Each request identifies the required task and the selected model. An intermediate orchestration layer validates the request and routes it to the corresponding speech-recognition or language-model service. This layer is implemented as a Python FastAPI service, with Pydantic schemas for request validation, configuration-driven task routing, the OpenAI Python client for compatible model endpoints and lightweight adapter classes that wrap each ASR or LLM backend behind a common interface. The orchestrator uses OpenAI-compatible interfaces when they are supported by the target service, while the Android application remains decoupled from these service-specific endpoints.

This separation allows registered models to be selected or replaced at the backend level without modifying the Android audio-capture or interaction logic. A new model can be incorporated by implementing an adapter in the orchestration layer. In the current default configuration, \texttt{\footnotesize whisper-large-v3-turbo}~\cite{radford2023whisper} is used for short-query transcription, WhisperX~\cite{bain2023whisperx} for conversation transcription and diarisation, and \texttt{\footnotesize gemma-4-31B} for answer generation. The embedding model, currently \texttt{\footnotesize all-MiniLM-L6-v2}, is configured independently by the conversational-memory service and associated with each vector index. Alternative language models, including \texttt{\footnotesize Mistral-Small-3.1-24B-Instruct-2503}, \texttt{\footnotesize EuroLLM-9B-Instruct} and \texttt{\footnotesize Apertus-8B-Instruct-2509}, as well as alternative ASR backends such as \texttt{\footnotesize MEUSLI-SLAM-ASR} and \texttt{\footnotesize Voxtral-Mini-3B-2507}, have also been deployed, registered with the orchestration layer, and successfully used within the assistant.

\subsection{On-device audio interaction}
The on-device component is implemented in Kotlin using native Android and Jetpack libraries. A persistent Android foreground \texttt{\footnotesize Service} manages the audio workflow, while the \texttt{ \footnotesize AudioRecord} API captures 16-bit PCM samples and writes them to a thread-safe circular buffer. A background predictor extracts and normalises MFCC features from the most recent audio window and feeds them to a lightweight GRU-based wake-word detector configured for the activation phrase ``OK Aura'', following the on-device architecture described in~\cite{lopez2023robustwakeupworddetection}. The model is packaged in TorchScript format and executed locally on the tablet. Audio capture and inference run outside the main thread to avoid blocking the Jetpack Compose interface. Android broadcast intents propagate recording states, audio levels, transcriptions, answers and source information to the application's ViewModel, which updates the interface asynchronously.

After wake-word activation, Silero VAD~\cite{silero_vad}
analyses incoming audio frames on the tablet to estimate
speech activity. Query capture uses a speech-probability
threshold of $0.5$ and a silence interval of $2$ seconds to determine when the utterance has ended. The resulting utterance, captured in PCM samples, is serialised as WAV audio and sent to the selected ASR service. The returned transcription becomes the retrieval query. Long-form conversation capture is handled by a separate recorder that continues acquiring audio while completed, user-configurable chunks are saved and uploaded asynchronously. 

\subsection{Turn-aware conversational memory}
After transcription and diarisation, the remote backend transforms timestamped, diarised segments into structured conversational units. Segment timestamps are first converted from chunk-relative positions to a shared conversation timeline. Given the chunk offset $t_c$ supplied by the Android client and local segment bounds $s_i$ and $e_i$, the conversation-level timestamps are
\vspace{-2mm}
\begin{equation}
t_i^{\mathrm{start}}=t_c+s_i,\qquad
t_i^{\mathrm{end}}=t_c+e_i.
\end{equation}

Consecutive segments are merged into turns only when both have valid timestamps, share the same known speaker label, and the gap between the preceding segment's end and the following segment's start lies between $-0.2$ and $1.0$ seconds. This allows a small timestamp overlap or a short pause. Otherwise, a new turn is started.

Because diarisation is performed independently for each audio chunk, the same participant may receive different speaker labels in successive chunks. Before passage construction, a speaker-reconciliation stage uses a language model to map these chunk-local labels to conversation-level speaker identifiers. The model receives the current turns, recent conversational context and a persistent speaker registry, and uses cues such as self-identification, direct address and the sequence of exchanges to propose speaker associations. This stage can also split a merged turn at preserved segment boundaries when the model identifies a likely speaker change. The registry is updated after successful ingestion and reused for subsequent chunks. This process aims to maintain consistent speaker identities, although its accuracy depends on the available textual context and the language model's assignments.

The resulting turns are grouped into passages using a
configurable target length in characters. Turns are added
to the current passage until the next turn would exceed
the target, at which point a new passage begins. Turns
are not split, so an individual turn may exceed the target.
Each passage retains speaker-labelled text, the start time
of its first turn, the end time of its last turn and its
turn count. However, fixed-duration audio chunks may cut through an ongoing spoken turn, and turns spanning successive chunks are not rejoined. The resulting passages and their metadata are passed to the embedding and vector-index update stage. For general document ingestion without structured turn information, the backend supports the configured text-splitting strategy, preserving compatibility with conventional document-based RAG workflows.

\subsection{Vector store, retrieval and response}
LanceDB is a vector database that stores embeddings alongside text and metadata and supports similarity-based retrieval. Each turn-aware passage is encoded with a configurable sentence-embedding model and stored with its text and metadata in a conversation-specific LanceDB table. The current default embedder is \texttt{\footnotesize all-MiniLM-L6-v2}, following the dense semantic-retrieval approach introduced by Sentence-BERT~\cite{reimers2019sbert}.
The embedding configuration is associated with each table and reused when new passages are appended, ensuring that stored passages and subsequent queries use the same representation space. After each successful chunk ingestion, new passages become searchable while recording continues.

At query time, the transcribed utterance is encoded using the embedding model associated with the selected conversation. LanceDB selects up to $k$ nearest passages. The retriever optionally filters them by a distance threshold and restores their original ingestion order to preserve the conversational sequence. The passage texts are included in the language-model context together with the user query, grounding generation in evidence from the selected conversation rather than solely in the model's parametric knowledge.

Grounding can improve contextual relevance but does not guarantee factual correctness, since retrieval, generation or the recorded source may still contain errors. Passage metadata, including timestamps and turn counts, is returned separately to the tablet for source display and excluded from the language-model prompt. Switching conversations changes the LanceDB table searched without requiring the pipeline to be rebuilt.

\section{Ethical considerations}
The current interface makes recording visible and provides explicit stop, deletion and source-inspection controls. Audio, transcripts and queries processed by external services are handled within the project infrastructure in accordance with the GDPR and applicable European data-protection requirements.

Should the system advance towards real-world deployment, it would additionally require informed consent from all participants, persistent recording notifications, private modes, clear data-retention policies and secure storage of conversation data. Future versions should also support user authentication, per-user permissions and appropriate speaker-verification mechanisms to prevent one resident from accessing another's sensitive information or mixing records belonging to different people. 

Finally, it should incorporate guardrails such as retrieval-confidence thresholds and consistency checks, communicate uncertainty and decline to answer when reliable evidence is unavailable. The assistant should always be presented as a recall aid rather than as an authoritative decision system.

\section{Conclusion}
We presented a voice-enabled household memory assistant that makes stored and recently captured conversations accessible through spoken queries. Its modular architecture separates the Android interaction layer, model services and conversational-memory pipeline, allowing compatible speech, embedding and language models to be selected or integrated without changing the assistant interaction logic. The turn-aware ingestion process preserves speaker and temporal context, while the interface supports conversation management and inspection of the timestamped passages used to support each answer.

\section{Acknowledgments}
This work has received funding from the European Union’s Horizon Europe research and innovation programme under the project ELOQUENCE (Grant Agreement No. 101135916).

\bibliographystyle{IEEEtran}
\bibliography{references}
\end{document}